\documentclass[acmjacm]{acmtrans2m}

\usepackage{amsmath,amsfonts}
\usepackage{bbm} 

\edef\qedrestoreat{\noexpand\catcode\lq\noexpand\@=\the\catcode\lq\@}
\catcode\lq\@=11

\ifx\protect\undefined\let\protect\relax\fi

\def\qed{\protect\@qed{$\qedsymbol$}}
\def\pushright{\protect\@pushright}

\def\QED{\protect\@qed{{\rm Q.E.D.}}}

\def\QEI{\protect\@qed{{\rm Q.E.I.}}}

\def\Proof{\protect\@Proof}\def\endProof{\protect\@endProof}%

\def\Proofof#1{\protect\@Proofof{#1}}\def\endProofof{\protect\@endProofof}%
\let\endproofof\endProofof

\def\qedsymbol{\raisebox{-.2ex}{$\Box$}}

\def\TheWordProof{\em Proof.}
\def\TheWordProofof#1{\em Proof of #1.}

\def\ProofFont{}

\newif\ifAutoQED\AutoQEDfalse

\newif\ifNumberResults
\ifx\theorem@style\undefined
   \NumberResultstrue
   
\fi

\def\parag@pushright#1{{
    \parfillskip=0pt            
    \widowpenalty=10000         
    \displaywidowpenalty=10000  
    \finalhyphendemerits=0      
    \hbox@pushright             
    #1
    \par}}

\def\hbox@pushright{
    \unskip                     
    \nobreak                    
    \hfil                       
    \penalty50                  
    \hskip.2em                  
    \null                       
    \hfill                      
}%

\newif\if@qed\@qedfalse
\def\save@set@qed{\let\saved@ifqed\if@qed\global\@qedtrue}%
\def\restore@qed{\global\let\if@qed\saved@ifqed}

\def\@Proof{%
   \par\removelastskip\bigskip\penalty100
   \save@set@qed
   \noindent\ProofFont{\TheWordProof\enskip}%
}%
\def\@Proofof#1{%
   \par\removelastskip\bigskip\penalty100
   \save@set@qed
   \noindent\ProofFont{\TheWordProofof{#1}\enskip}%
}%
\def\@endProof{%
   \qed\restore@qed
   \penalty-100 \medskip
}
\def\@endProofof{%
   \qed\restore@qed
   \penalty-100 \medskip
}
\def\@qed#1{%
\if@qed                                 
     \global\@qedfalse
        \ifmmode\ifinner\pushright{#1}
        \else\eqno{\qedsymbol}\fi
        \else\pushright{#1}\fi%
\else\ifhmode\ifinner\else\par\fi\fi
\fi}
\def\@pushright#1{%
  {\ifvmode                             
       \null\hfill{#1}\par              
  \else\ifmmode\maths@pushright{\hbox{#1}}
       \else\ifinner\hbox@pushright{#1}
            \else\parag@pushright{#1}
  \fi  \fi  \fi
}}%
\def\maths@pushright#1{{%
  \ifinner
     \hbox@pushright{#1}%
  \else
     \eqno#1
     \def\]{$$\ignorespaces}
  \fi
}}%
\newtheorem{theorem}{Theorem}[section]

\newtheorem{corollary}[theorem]{Corollary}
\newtheorem{proposition}[theorem]{Proposition}
\newtheorem{lemma}[theorem]{Lemma}
\newdef{definition}[theorem]{Definition}
\newdef{remark}[theorem]{Remark}

\newcommand{\C}{\mathbb{C}}
\newcommand{\N}{\mathbb{N}}

\newcommand{\R}{\mathbb{R}}

\newcommand{\mcal}[1]{\mathcal{#1}}

\newcommand{\Rig}[2]{\mathrm{rig}_{#1}(#2)}
\newcommand{\vol}[2]{\mathrm{vol}_{#1}(#2)}
\newcommand{\Var}[1]{\mathrm{Var}(#1)}
\newcommand{\avol}[1]{\mathrm{msv}_{#1}}

\newcommand{\rk}[1]{\mathrm{rk}(#1)}
\newcommand{\dist}[1]{\mathrm{dist}(#1)}
\newcommand{\size}[1]{\mathcal{S}(#1)}

\newcommand{\DFT}{\mathrm{DFT}_n}
\newcommand{\Cov}[1]{\mathrm{Cov}(#1)}
\newcommand{\compl}[1]{\mathcal{C}(#1)}
\newcommand{\diag}[1]{\mathrm{diag}(#1)}
\newcommand{\mean}[1]{\mathrm{E}[#1]}
\newcommand{\prob}[1]{\mathrm{P}\left[#1\right]}

\newcommand{\hcompl}[2]{\mathcal{C}_{#1}(#2)}
\newcommand{\Circ}[1]{\mathrm{Circ}(#1)}
\newcommand{\ew}[1]{\mathrm{E}(#1)}

\title{Lower Bounds on the Bounded Coefficient Complexity of Bilinear Maps}

\author{Peter B\"urgisser \\
        Universit\"at Paderborn
        \and Martin Lotz \\
        Universit\"at Paderborn}

\begin{abstract}
We prove lower bounds of order $n\log n$ for both the problem to multiply
polynomials of degree~$n$, and to divide polynomials with remainder, in the
model of bounded coefficient arithmetic circuits over the complex numbers.
These lower bounds are optimal up to order of magnitude.
The proof uses a recent idea of R.~Raz [Proc.\ 34th STOC 2002] proposed for
matrix multiplication.
It reduces the linear problem to multiply a random circulant matrix with a
vector to the  bilinear problem of cyclic convolution.
We treat the arising linear problem by extending J.~Morgenstern's bound
[J.\ ACM 20, pp.\ 305-306, 1973] in a unitarily invariant way. This
establishes a new lower bound on the bounded coefficient complexity of
linear forms in terms of the singular values of the corresponding matrix.
In addition, we extend these lower bounds for linear and bilinear maps to 
a model of circuits that allows a restricted number of unbounded scalar 
multiplications.
\end{abstract}

\category{F.1.1}{Computation by Abstract Devices}{Models of Computation}
\category{F.2.1}{Analysis of Algorithms and Problem Complexity}{Numerical Algorithms and Problems}
\category{I.1.2}{Symbolic and Algebraic Manipulation}{Algorithms}
\terms{Algorithms, Theory}
\keywords{algebraic complexity, bilinear circuits, lower bounds, singular values}

\begin{document} 

\begin{bottomstuff}
Author's address:  Faculty of Computer Science, Electrical Engineering, and Mathematics,
University of Paderborn, 33095 Paderborn, Germany. Email: \{pbuerg,lotzm\}@math.uni-paderborn.de.
\end{bottomstuff}

\maketitle

\section{Introduction}

Finding lower bounds on the complexity of polynomial functions over the complex
numbers is one of the fundamental problems of algebraic complexity theory.
It becomes more tractable if we restrict the model of computation to
arithmetic circuits, where the multiplication with scalars is restricted to
constants of bounded absolute value. This model was introduced in a seminal work 
by~\cite{morg:73,morg:75}, where it was 
proved that the complexity of multiplying a vector with some given square matrix~$A$
is bounded from below by the logarithm of the absolute value of the determinant of $A$.
As a consequence, Morgenstern derived the lower bound $\frac{1}{2}n\log{n}$ for
computing the Discrete Fourier Transform.

\cite{vali:76-2,vali:77} analyzed the problem to prove nonlinear lower bounds
on the complexity of the Discrete Fourier Transform and related linear problems in the
unrestricted model of arithmetic circuits. However, despite many attempts, this
problem is still open today.

To motivate the bounded coefficient model (b.c.\ for short),
we note that many algorithms for arithmetic problems, like the Fast Fourier Transform 
and the fast algorithms based on it, use only small constants.
\cite{chaz:98} advocated the b.c.\ model as a natural model of computation 
by arguing that the finite representation of numbers is essentially equivalent 
to bounded coefficients. 

\cite{chaz:98} refined Morgenstern's bound by proving a lower bound on the 
b.c.\ linear complexity of a matrix~$A$ in terms of the singular values of $A$.
His applications are nonlinear lower bounds for range searching problems.
Several papers~\cite{niwi:95,loka:95,pudl:98} provided size-depth trade-offs for b.c.\ arithmetic circuits.
The concept of matrix rigidity, originally introduced in~\cite{vali:77}, 
hereby plays a vital role. 
A geometric variant of this concept (euclidean metric instead of Hamming metric) 
is closely related to the singular value decomposition of a matrix and turns out to be
an important tool, as worked out in~\cite{loka:95}. 
\cite{rraz:02} recently proved a nonlinear lower bound on the complexity of
matrix multiplication in the b.c.\ model. To our knowledge, this paper and~\cite{niwi:95} 
are the only ones which deal with 
the complexity of bilinear maps in the b.c.\ model of computation.

The main result of this paper (Theorem~\ref{main}) 
is a nonlinear lower bound of order $n\log n$
to compute the cyclic convolution of two given vectors in the b.c.\ model.
This bound is optimal up to a constant factor.
The proof is based on ideas in~\cite{rraz:02} to establish a lower bound on the
complexity of a bilinear map $(x,y)\mapsto\varphi(x,y)$ in terms of the complexity of the linear maps
$y\mapsto\varphi(a,y)$ obtained by fixing the first input to $a$
(Lemma~\ref{lemma21}). However,
the linear circuit for the computation of $y\mapsto\varphi(a,y)$ resulting from a hypothetical
b.c.\ circuit for $\varphi$ has to be transformed into a \textit{small}
one with bounded coefficients.
This can be achieved with a geometric rigidity argument by choosing a vector~$a$
at random according to the standard normal distribution
in a suitable linear subspace of $\C^m$ (Lemma~\ref{lemma1}). 

In the case of matrix multiplication, \cite{rraz:02} proceeded by applying 
a geometric rigidity bound
to the resulting linear problem via the Hoffman-Wielandt inequality. 
This approach does not yield good enough bounds in our situation,
where we have to estimate the complexity of structured random matrices; in the case of
the convolution these are circulant matrices.
Instead, we treat the arising linear problem by extending Morgenstern's bound
in a new way. We define the $r$-{\em mean square volume} of a complex matrix~$A$, 
which turns out to be the square root of the $r$-th elementary symmetric function 
in the squares of the singular values of~$A$.
An important property of this quantity is that it is invariant under multiplication with
unitary matrices from the left or the right. We prove that the logarithm of the
$r$-mean square volume provides a lower bound on the b.c.\ complexity of the matrix~$A$
(Proposition~(\ref{pro:msv-bound})). 
This implies that the logarithm of the product 
of the largest $r$ singular values is a lower bound on the b.c.\ complexity.

We also study an extension of the bounded coefficient model of computation by allowing 
a limited number of {\em help gates} 
corresponding to scalar multiplications with unbounded constants. 
We can show that our proof technique is robust in the sense that it still allows to prove 
$n\log{n}$ lower bounds if the number of help gates is restricted to 
$(1-\epsilon)n$ for fixed $\epsilon>0$. 
This is achieved by an extension of the mean square volume bound (Proposition~\ref{th:msv-help}), 
which is related to the spectral lemma in~\cite{chaz:98}. 
The proof is based on some matrix perturbation arguments.  

From the lower bound for the cyclic convolution we obtain nonlinear lower bounds for
polynomial multiplication, inversion of power series, and polynomial division with remainder
by noting that the well-known reductions between these problems~\cite{bucs:96}
preserve the b.c.\ property. These lower bounds are again optimal up to order of magnitude.

\subsection{Organization of the paper}

In Section~\ref{se:model}, we introduce the model of computation and discuss known facts 
facts about singular values and matrix rigidity.
We also introduce some notation and present auxiliary results related to 
(complex) Gaussian random vectors. 
In Section~\ref{se:msv-bound} we first recall previously known lower bounds 
for b.c.\ linear circuits.
Then we introduce the mean square volume of a matrix and prove an extension of 
Morgenstern's bound in terms of this quantity.
Section~\ref{se:conv} contains the statement and proof of our main
theorem, the lower bound on cyclic convolution.
In Section~\ref{se:appl}, we derive lower bounds for polynomial multiplication,
inversion of power series and division with remainder.
Finally, in Section~\ref{se:help} we show that our results can be extended to the case, where 
a limited number of unbounded scalar multiplications (help gates) is allowed.  

\section{Preliminaries}
\label{se:model}

We start this section by giving a short introduction to the model of
computation. 

\subsection{The model of computation}

We will base our arguments on the model of algebraic straight-line
programs over~$\C$, which are often called arithmetic circuits in the literature.
For details on this model we refer to chapter 4 of \cite{bucs:96}.
By a result in~\cite{stra:73-1}, we may exclude divisions without 
loss of generality. 

\begin{definition}
A \textit{straight-line program}~$\Gamma$ expecting inputs of length $n$ is a
sequence $(\Gamma_1,\dots,\Gamma_r)$ of instructions
$\Gamma_s=(\omega_s;i_s,j_s)$, $\omega_s \in \{*,+,-\}$ or
$\Gamma_s=(\omega_s;i_s)$, $\omega_s\in \C$, with 
integers $i_s, j_s$ satisfying $-n<i_s,j_s<s$. A sequence of
polynomials $b_{-n+1},\dots,b_r$ is called the \textit{result sequence}
of $\Gamma$ on input variables $a_1,\dots,a_n$, if for $-n<s\leq 0$,
$b_s=a_{n+s}$, and for $1\leq s\leq r$, $b_s=b_{i_s}\omega_sb_{j_s}$ if
$\Gamma_s=(\omega_s;i_s,j_s)$ and $b_s=\omega_sb_{i_s}$ if
$\Gamma_s=(\omega_s;i_s)$. $\Gamma$ is said to
\textit{compute} a set of polynomials $F$ on input $a_1,\dots,a_n$, if
the elements in $F$ are among those of the result sequence of $\Gamma$ on
that input. The \textit{size} $\size{\Gamma}$ of $\Gamma$ is the number~$r$ 
of its instructions.    
\end{definition}

In the sequel we will refer to such straight-line programs briefly as
circuits. 
A circuit in which the scalar multiplication is restricted
to scalars of absolute value at most $2$ will be called a
\textit{bounded coefficient circuit} (b.c.\ circuit for short).
Of course, the bound of~$2$ could be replaced by any other fixed bound. 
Any circuit can be transformed into a b.c.\ circuit by replacing a
multiplication with a scalar $\lambda$ with at most
$\log{|\lambda|}$ additions and a multiplication with a
scalar of absolute value at most~$2$. 
Unless otherwise stated, $\log$ will always refer to 
logarithms to the base~$2$.

We now introduce restricted notions of circuits, designed for
computing linear and bilinear maps. 

\begin{definition}
\label{def:circuit}
A circuit $\Gamma=(\Gamma_1,\dots,\Gamma_r)$ expecting inputs $X_1,\dots,X_n$ is called a
\textit{linear circuit}, if $\omega_s\in \{+,-\}$ for every
instruction $\Gamma_s=(\omega_s;i_s,j_s)$, or $\omega_s\in\C$ if the
instruction is of the form $(\omega_s;i_s)$. 
A circuit on inputs $X_1,\dots,X_m,Y_1,\dots,Y_n$ is called a
\textit{bilinear circuit}, if its sequence of instructions can be
partitioned as $\Gamma=(\Gamma^{(1)},\Gamma^{(2)},\Gamma^{(3)},\Gamma^{(4)})$, where
\begin{enumerate}
\item $\Gamma^{(1)}$ is a linear circuit with the $X_i$ as inputs,
\item $\Gamma^{(2)}$ is a linear circuit with the $Y_j$ as inputs,
\item each instruction from $\Gamma^{(3)}$ has the form $(*;i,j)$,
with $\Gamma_i \in \Gamma^{(1)}$ and $\Gamma_j \in \Gamma^{(2)}$,
\item $\Gamma^{(4)}$ is a linear circuit with the previously
computed results of $\Gamma^{(3)}$ as inputs.
\end{enumerate}
In other words, $\Gamma^{(1)}$ and $\Gamma^{(2)}$ compute
linear functions $f_1,\dots,f_k$ in the $X_i$ and $g_1,\dots,g_\ell$ in the $Y_j$.  
$\Gamma^{(3)}$ then multiplies the $f_i$ with the $g_j$ and
$\Gamma^{(4)}$ computes linear combinations of the products $f_i g_j$. 
\end{definition}

It is clear that linear circuits compute linear maps and that bilinear
circuits compute bilinear maps. On the other hand, it can be shown
that any linear (bilinear) map can be computed by a linear (bilinear)
circuit such that the size increases at most by a constant factor
(cf.\ \cite[Theorem~13.1, Proposition~14.1]{bucs:96}).  
This remains true when considering bounded coefficient
circuits, as can easily be checked. 
From now on, we will only be concerned with bounded coefficient circuits.

\begin{definition}
By the {\em b.c.\ complexity} $\compl{\varphi}$ of a bilinear map 
$\varphi \colon \C^m\times \C^n \rightarrow \C^p$ we understand the size 
of a smallest b.c.\ bilinear circuit computing $\varphi$.
By the {\em b.c.\ complexity} $\compl{\varphi^A}$ of a linear map 
$\varphi^A\colon\C^n\rightarrow \C^m$ (or the corresponding matrix $A\in\C^{m\times n}$),
we understand the size of a smallest b.c.\ linear circuit computing $\varphi^A$.
\end{definition}

By abuse of notation, we also write $\compl{F}$ for the smallest size of a b.c.\ circuit 
computing a set $F$ of polynomials from the variables. 
(There is no serious danger of confusion arising from this, since these complexity notions differ 
at most by a constant factor.) 

Let $\varphi\colon \C^m\times \C^n \rightarrow \C^p$ be a bilinear map 
described by $\varphi_k(X,Y)=\sum_{i,j} a_{ijk}X_iY_j$. Assuming $|a_{ijk}|\leq 2$,
it is clear that $\compl{\varphi}\leq 3mnp$. Therefore, if
$f_1,\dots,f_k$ are the linear maps computed on the first set of
inputs by an optimal b.c.\ bilinear circuit for $\varphi$, 
we have $k\leq \size{\Gamma} \leq 3mnp$.
 
The complexity of a bilinear map $\varphi$ can be related to the complexity of the
associated linear map $\varphi(a,-)$, where $a\in\C^m$. 
We have taken the idea behind the following lemma from~\cite{rraz:02}.

\begin{lemma}
\label{lemma21}
Let $\varphi\colon \C^m\times \C^n\rightarrow \C^p$ be a bilinear map and
$\Gamma$ be a b.c.\ bilinear circuit
computing $\varphi$. If $f_1,\dots,f_k$ are the linear maps computed
by the circuit on the first set of inputs, then for
all $a\in \C^m$: 
\begin{equation*}
  \compl{\varphi(a,-)}\leq \size{\Gamma}+p\log{(\max_j{|f_j(a)|})}.
\end{equation*}
\end{lemma}

\begin{Proof}
Let $a\in \C^m$ be chosen and set  $\gamma=\max_j{|f_j(a)|}$. 
Transform the circuit $\Gamma$ into a linear circuit $\Gamma'$ by the following steps:
\begin{enumerate}
\item replace the first argument $x$ of the input by $a$,
\item replace each multiplication by $f_i(a)$ with a multiplication
by $2\gamma^{-1}f_i(a)$,
\item multiply each output with $\gamma/2$ by simulating this with at
most $\log{(\gamma/2)}$ additions and one multiplication with a scalar
of absolute value at most $2$. 
\end{enumerate}
This is a b.c.\ linear circuit computing the map
$\varphi(a,-)\colon~\C^n\rightarrow \C^p$. Since there are $p$
outputs, the size increases by at most $p\log{\gamma}$.
\end{Proof}

\subsection{Singular values and matrix rigidity}

The \textit{Singular Value Decomposition} (SVD) is one of the most important
matrix decompositions in numerical analysis. Lately, it has also come to play a
prominent role in proving lower bounds for linear circuits~\cite{chaz:98,loka:95,rraz:02}. 
In this section, we present some basic facts about singular values and show how they relate
to notions of matrix rigidity. For a more detailed account on the SVD, we refer to~\cite{govl:96}. 
We also find~\cite[Chapt.~1, Sect.~4]{couh:31} a useful reference. 

The singular values of $A\in \C^{m\times n}$,
$\sigma_1\ge \ldots \ge\sigma_{\min\{m,n\}}$,  
can be defined as the square roots of the 
eigenvalues of the hermitian matrix $AA^*$. 
Alternatively, they can be characterized as follows: 
\begin{equation*}\label{eq:svdist}
  \sigma_{r+1} = \min\{ \|A-B\|_2 \mid B\in\C^{m\times n},\rk{B}\le r\},
\end{equation*}
where $\|\cdot\|$ denotes the matrix $2$-norm. 
An important consequence is the Courant-Fischer min-max theorem stating 
\begin{equation*}\label{eq:cf}
  \sigma_{r+1}\ =\min_{\mathrm{codim}V = r}\ \max_{x\in V-\{0\}}\frac{\|Ax\|_2}{\|x\|_2}.
\end{equation*}
This description implies the following useful fact from matrix perturbation theory:  
\begin{equation}\label{eq:pert}
\sigma_{r+h}(A) \le \sigma_r(A + E)
\end{equation}
if the matrix $E$ has rank at most~$h$.
 
More generally, for any metric d on $\C^{m\times n}$ (or $\R^{m\times n}$) and $1\leq r\leq \min{\{m,n\}}$, 
we can define the $r$-\textit{rigidity} of a matrix $A$ to be the distance of $A$ to the
set of all matrices of rank at most $r$ with respect to this metric:
\begin{equation*}
  \Rig{\mathrm{d},r}{A}=\min\{ \mathrm{d}(A,B) \mid B\in\C^{m\times n},\rk{B}\le r\}.
\end{equation*}
Using the Hamming metric, we obtain the usual matrix rigidity as introduced in~\cite{vali:77}.
On the other hand, using the metric induced by the $1,2$-norm $\|A\|_{1,2}:=\max_{\|x\|_1=1}{\|Ax\|_2}$,
we obtain the following geometric notion of rigidity, as introduced in~\cite{rraz:02}:
\begin{equation*}\label{eq:razrig}
   \Rig{r}{A}\ =\min_{\dim{V}=r}\ \max_{1\leq i\leq n}\text{dist}(a_i,V).
\end{equation*}
Here, the $a_i$ are the column vectors of $A\in \C^{m\times n}$ and
dist denotes the usual euclidean distance. 

Notions of rigidity can be related to one another the same way the underlying norms can. In
particular, we have the following relationship between the geometric rigidity and the singular
values:
$$
 \frac{1}{\sqrt{n}}\, \sigma_{r+1}(A) \le \Rig{r}{A} \le \sigma_{r+1}(A).
$$
The proofs of these inequalities are based on well known inequalities for matrix norms. 
For instance, if $B$ is a matrix of rank at most~$r$ with columns $b_i$, we have 
$$
 ||A -B||^2_{1,2} = \max_i ||a_i-b_i||_2^2 \ge \frac{1}{n}\sum_{i=1}^n ||a_i-b_i||^2_2
  \ge \frac{1}{n} ||A-B||^2_2 \ge \frac{1}{n}\sigma_{r+1}^2 ,
$$
which shows the left inequality. 

\subsection{Complex Gaussian vectors}\label{sse:crv}

A random vector $X=(X_1,\ldots,X_n)$ in $\R^n$ is called 
{\em standard Gaussian}
iff its components $X_i$ are i.i.d.\ standard normal distributed. 
It is clear that an orthogonal transformation of such a random vector 
is again standard Gaussian. 

Throughout this paper, we will be working with random vectors $Z$ assuming values in $\C^n$. 
However, by identifying $\C^n$ with $\R^{2n}$, we can think of $Z$ as a $2n$-dimensional real random vector. 
In particular, it makes sense to say that such~$Z$ is (standard) Gaussian in $\C^n$. 

Let $U$ be an $r$-dimensional linear subspace of $\C^n$. 
We say that a random vector $Z$ with values in $U$ 
is {\em standard Gaussian in $U$} iff for some 
orthonormal basis $b_1,\dots,b_r$ of $U$ we have 
$Z=\sum_j \zeta_j b_j$, where the random vector $(\zeta_j)$ of the components is 
standard Gaussian in $\C^r$.
It is easy to see that this description does not depend on the choice of the orthonormal basis. 
In fact, the transformation of a standard Gaussian vector with a unitary matrix 
is again standard Gaussian, since a unitary transformation $\C^r\to\C^r$ induces 
an orthogonal transformation $\R^{2r}\to\R^{2r}$. 

The easy proof of the following lemma is left to the reader. 

\begin{lemma}\label{le:expd}
Let $(Z_1,\ldots,Z_n)$ be standard Gaussian in $\C^n$.
Consider a complex linear combination $S=f_1 Z_1+ \ldots + f_n Z_n$ with 
$f=(f_1,\ldots,f_n)\in\C^n$. 
Then the real and imaginary parts of $S$ are
independent and normal distributed, each with mean $0$ and variance $\|f\|^2$. 
Moreover, $T:=|S|^2/{2 \|f\|^2}$ is exponentially distributed with parameter~$1$. That is, 
the density function is $e^{-t}$ for $t\geq 0$ and  
the mean and the variance of~ $U$ are both equal to~$1$. 
\end{lemma}

\subsection{Two useful inequalities}
\label{se:useful-ineq}

Let $X,Y$ be i.i.d.\ standard normal random variables and set
$\gamma:=1-\mean{\log{X^2}}$ and $\theta:=\mean{\log^2 (X^2+Y^2)}$. 
Evaluating the corresponding integrals yields
\begin{align*}
  \gamma&=-\frac{1}{\sqrt{\pi}}\int_0^{\infty} t^{-1/2} e^{-t} \log{t}\, dt \approx 2.83\\
  \theta&=\frac{1}{2}\int_0^{\infty} e^{-t/2} \log^2{t}\, dt \approx 3.45.
\end{align*}

\begin{lemma}
\label{prop22}
Let $Z$ be a centered Gaussian variable with complex values. Then
$$
  0 \le \log{\mean{|Z|^2}} - \mean{\log{|Z|^2}} \leq \gamma, \  \Var{\log{|Z|^2}} \leq \theta.
$$
\end{lemma}

\begin{Proof}
By a principal axis transformation, we may assume that 
$Z=\lambda_1 X+i\lambda_2 Y$ with independent standard normal $X,Y$. 
The difference 
$\Delta:= \log{\mean{|Z|^2}}-\mean{\log{|Z|^2}}$
is nonnegative, since $\log$ is concave (Jensen's inequality). 
By linearity of the mean, $\Delta$ as well as $\Var{\log{|Z|^2}}$ 
are invariant under multiplication of $Z$ with scalars. We may
therefore w.l.o.g.\ assume that $1=\lambda_1\geq \lambda_2$. From
this we see that
\begin{align*}
  \log{\mean{|Z|^2}}&=\log{\mean{X^2+\lambda_2^2Y^2}}\leq \log{\mean{X^2+Y^2}}=1\\
  \mean{\log{|Z|^2}}&=\mean{\log{(X^2+\lambda_2^2Y^2)}}\geq \mean{\log{X^2}}=1-\gamma,
\end{align*}
which implies the first claim. 
The estimates 
$$
 \Var{\log{|Z|^2}} \le \mean{\log^2{|Z|^2}} \le \mean{\log^2 (X^2+Y^2)} = \theta .
$$
prove the second claim. 
\end{Proof}

\section{The mean square volume bound}
\label{se:msv-bound}

Morgenstern's bound~\cite{morg:73} states that 
$\compl{A} \geq \log{|\det{(A)}|}$ for a square matrix~$A$,
see also~\cite[Chapter~13]{bucs:96} for details. 
We are going to study several generalizations of this bound. 

Let $A\in \C^{m\times n}$ be a matrix. 
For an $r$-subset $I\subseteq [m]:=\{1,\dots,m\}$ let $A_I$ denote the submatrix of $A$
consisting of the rows indexed by $I$. 
The Gramian determinant $\det A_I A_I^*$ can be interpreted as the square of the 
volume of the parallelepiped spanned by the rows of $A_I$
($A^*$ denotes the complex transpose of $A$).

\cite{rraz:02} defined the $r$-\textit{volume} of $A$ by 
\begin{equation*}\label{def:r-vol}
 \vol{r}{A} := \max_{|I|=r}\ (\det A_I A_I^*)^{1/2} 
\end{equation*}
and observed that the proof of Morgenstern's bound extends to 
the following {\em $r$-volume bound}: 
\begin{equation}\label{eq:r-vol-bound}
 \compl{A} \ge \log{\vol{r}{A}}. 
\end{equation}
Moreover, \cite{rraz:02} related this quantity to the geometric rigidity as follows:
\begin{equation*}\label{eq:r-vol-rig}
 \vol{r}{A} \ge (\Rig{r}{A})^r ,
\end{equation*}
which implies the {\em rigidity bound}, 
\begin{equation}\label{eq:rig-bound}
 \compl{A} \ge r \log\Rig{r}{A} .
\end{equation}

For our purposes it will be important to work with a variant of the $r$-volume 
that is completely invariant under unitary transformations. 
Instead of taking the maximum of the volumes $(\det A_I A_I^*)^{1/2}$,  
we will use the sum of the squares. 
We define the $r$-\textit{mean square volume} $\avol{r}(A)$ of $A\in \C^{m\times n}$ by 
\begin{equation*}\label{def:msv}
  \avol{r}(A) := \bigg(\sum_{|I|=r} \det A_I A_I^*\bigg)^{1/2} 
               = \bigg(\sum_{|I|=|J|=r} |\det{A_{I,J}}|^2\bigg)^{1/2}. 
\end{equation*}
Hereby, $A_{I,J}$ denotes the $r\times r$ submatrix consisting of the rows
indexed by $I$ and columns indexed by $J$. The second equality is a 
consequence of the Binet-Cauchy formula
$\det{A_IA_I^{*}} = \sum_{|J|=r}|\det{A_{I,J}}|^2$, 
see~\cite[Chapter~4]{bell:97}.   
The choice of the $L_2$-norm instead of the maximum norm results in the 
following inequality
\begin{equation}\label{eq:cn}
  \vol{r}{A}\leq \avol{r}(A)\leq \sqrt{\binom{m}{r}}\ \vol{r}{A} .
\end{equation}
The mean square volume has the following nice properties, 
which are all easy to verify:
$$
\avol{r}(A)=\avol{r}(A^*),\ \avol{r}(\lambda A)=~|\lambda|^r\, \avol{r}(A), 
 \ \avol{r}(A)=\avol{r}(UAV),
$$ 
where $\lambda\in\C$ and $U$ and $V$ are unitary matrices of the correct format. 
Note also that $\avol{n}(A)=|\det A|$ for $A\in\C^{n\times n}$. 
The unitary invariance allows to express the mean square volume of $A$ in terms of 
the  singular values $\sigma_1\ge\ldots\dots\ge\sigma_p$ of~$A$, $p:=\min{\{m,n\}}$.
It is well known~\cite{govl:96} that there are unitary matrices
$U\in \C^{m\times m}$ and $V\in \C^{n\times n}$ such that
$U^{*}AV=\diag{\sigma_1,\dots,\sigma_p}$. Hence we obtain 
\begin{equation}
\label{eq:prop31}
  \avol{r}^2(A)=\avol{r}^2(\diag{\sigma_1,\dots,\sigma_p})=\ \sum_{|I|=r}\prod_{i\in I}\sigma_{i}^2 
            \ \ge\ \sigma_1^2 \sigma_2^2 \cdots \sigma_r^2,
\end{equation}
where $I$ runs over all $r$-subsets of $[p]$. 
Hence, the square of the $r$-mean square volume of a matrix is the $r$-th
elementary symmetric polynomial in the squares of its singular values.  

Combining the $r$-volume bound~(\ref{eq:r-vol-bound}) 
with (\ref{eq:cn}) we obtain 
the following {\em mean square volume bound}.

\begin{proposition}\label{pro:msv-bound}
For a matrix $A\in \C^{m\times n}$ and $r\in\N$ with $1\le r\le \min\{m,n\}$ 
we have
\begin{equation}\label{eq:msv-bound}
  \compl{A}\geq \log{\avol{r}(A)}-\frac{m}{2}.
\end{equation}
\end{proposition}

\begin{remark}
The $r$-volume can be seen as the $1,2$-norm of the
map $\Lambda^rA$ induced by $A$ between the exterior algebras
$\Lambda^r\C^n$ and $\Lambda^r\C^m$ (see e.g., \cite{lang:84} 
for background on multilinear algebra). 
Similarly, the mean square volume can be interpreted as the Frobenius norm
of $\Lambda^rA$. 
The unitary invariance of the mean square volume 
also follows from the fact that $\Lambda^r$ is equivariant with respect to 
unitary transformations and that the Frobenius norm is invariant under such. 
\end{remark}

\section{A lower bound on cyclic convolution}
\label{se:conv}

In this section we use the mean square volume bound~(\ref{eq:msv-bound}) 
to prove a lower bound on the bilinear map of the cyclic convolution. 

Let $f=\sum_{i=0}^{n-1}a_ix^i$ and $g=\sum_{i=0}^{n-1}b_ix^i$ be polynomials
in $\C[X]$. The cyclic convolution of $f$ and $g$ is the polynomial 
$h=\sum_{i=0}^{n-1}c_i x^i$, 
which is given by the product of $f$ and $g$ in the
quotient ring $\C[X]/(X^n-1)$. More explicitly:
\begin{equation*}
  c_k=\sum_{i+j\equiv k \bmod n} a_i b_j, \quad 0\leq k < n .
\end{equation*}
Cyclic convolution is a bilinear
map on the coefficients. For a fixed polynomial with coefficient
vector $a=(a_0,\dots,a_{n-1})$, this map turns into a linear
transformation with the circulant matrix
\begin{equation*}
  \Circ{a} = \begin{pmatrix} a_0 & a_1 & \dots & a_{n-1}\\
                    a_{n-1} & a_0 & \dots & a_{n-2}\\
                    \dots & \dots & \dots & \dots\\
                    a_1 & a_2 & \dots & a_0
   \end{pmatrix}.
\end{equation*}

Let $\DFT=(\omega^{jk})_{0\leq j,k<n}$ be the
matrix of the Discrete Fourier Transform, with $\omega=e^{2\pi i/n}$. 
It is well known~\cite[Sect.~4.7.7]{govl:96} that  
\begin{equation*}
\label{dft}
  \Circ{a} = \big(\frac{1}{\sqrt{n}}\DFT\big)^{-1}\diag{\lambda_0,\dots,\lambda_{n-1}}\frac{1}{\sqrt{n}}\DFT,
\end{equation*}
where the eigenvalues $\lambda_k$ of $\Circ{a}$ are given by
\begin{equation}\label{eq:ev-DFT}
  (\lambda_0,\dots,\lambda_{n-1})^{\top}=\DFT(a_0,\dots,a_{n-1})^{\top}.
\end{equation}
Hence the singular values of $\Circ{a}$ are 
$|\lambda_0|,\ldots,|\lambda_{n-1}|$ (in some order).  
Note that $n^{-1/2}\DFT$ is unitary. 

We recall that the Fast Fourier Transform provides a 
b.c.\ bilinear circuit of size $O(n\log{n})$ that computes
the $n$-dimensional cyclic convolution. 
The main result of the paper is the optimality of this 
algorithm in the b.c.\ model.

\begin{theorem}
\label{main}
The bounded coefficient complexity of the $n$-dimensional cyclic convolution 
$\mathrm{conv}_n$ satisfies 
$\compl{\mathrm{conv}_n} \geq \frac{1}{12}n\log{n}-O(n\log\log{n})$.
\end{theorem}

In fact, the proof of the theorem shows that we can replace the constant factor $1/12$ 
by the slightly larger value~$0.086$. We state the theorem with $1/12$ for 
simplicity of exposition. 

\subsection{Bounding the absolute values of linear forms}

To prepare for the proof, we need some lemmas. 
The idea behind the following lemma is already present in \cite{rraz:02}.
We will identify linear forms on $\C^n$ with vectors in $\C^n$. 

\begin{lemma}
\label{lemma1}
Let $f_1,\dots,f_k \in \C^n$ be linear forms and let $1\leq r<n$.
Then there exists a complex subspace $U\subseteq \C^n$ of dimension $r$
such that for a standard Gaussian vector in $U$, we have
\begin{equation*}
  \prob{\max_i{|f_i(a)|}\leq 2\sqrt{\ln{(4k)}}\ \Rig{n-r}{f_1,\dots,f_k}}\geq \frac{1}{2}.
\end{equation*}
\end{lemma}
 
\begin{Proof}
Set $R=\Rig{n-r}{f_1,\dots,f_k}$. Then there exists a linear subspace
$V \subseteq \C^n$ of dimension $n-r$ such that 
$\dist{f_i,V}\leq R$ for all $1\leq i\leq k$.  
Let $f_i'$ be the projection of $f_i$ along $V$ onto the orthogonal complement 
$U:=V^{\bot}$ of~$V$. By our choice of the subspace $V$ we have $\|f_i'\|\leq R$. 

Let $(b_1,\dots,b_n)$ be standard Gaussian in $\C^n$ 
and $a$ be the orthogonal projection of $b$ onto $U$ along~$V$. 
Then $a$ is standard Gaussian in $U$. 
Moreover, we have $f'_i(b)=f_i(a)$. 
By Lemma~\ref{le:expd}, the random variable 
$T=|f_i'(b)|^2/(2\|f_i'\|^2)$ is exponentially distributed with
parameter $1$. 

The assertion now follows from standard large deviations arguments. 
For any real $\lambda$, we have
\begin{equation*}
  \prob{T\geq\lambda}=\mean{1_{T\geq \lambda}}\leq \mean{e^{(T-\lambda)/2}}=e^{-\lambda/2}\mean{e^{T/2}}.
\end{equation*}
On the other hand,
\begin{equation*}
  \mean{e^{T/2}}=\sum_{k=0}^{\infty}\frac{1}{2^kk!}\mean{T^k}=\sum_{k=0}^{\infty}\frac{1}{2^k}=2,
\end{equation*}
since $\mean{T^k}=\int_0^{\infty}x^ke^{-x}dx=k!$.
It follows that
\begin{align*}
  \prob{T\geq\lambda}&=\prob{|f_i'(b)|^2\geq 2\lambda\|f_i'\|^2}\leq2e^{-\lambda/2}.
\end{align*}
Since $\|f_i'\|\leq R$, we have for a fixed $i$ that
\begin{equation*}
  \prob{|f_i(a)|\geq\sqrt{2\lambda}\, R}\leq 2e^{-\lambda/2} .
\end{equation*}
By the union bound we obtain
\begin{equation*}
  \prob{\max_i|f_i(a)|\geq\sqrt{2\lambda}\, R}\leq 2ke^{-\lambda/2}.
\end{equation*}
Setting $\lambda=2\ln{(4k)}$ completes the proof.
\end{Proof}

\subsection{Proof of the main result}

In the next lemma, we state a lower bound on the b.c.\ linear
complexity of the circulant. 

\begin{lemma}
\label{lemma2}
Let $U\subseteq \C^n$ be a subspace of dimension $r$. 
For a standard Gaussian vector~$a$ in $U$, we have
\begin{equation*}
  \prob{\compl{\Circ{a}}\geq \frac{1}{2}r\log{n} - c n}>\frac{1}{2},
\end{equation*}
where $c=\frac{1}{2}(2 + \gamma+\sqrt{2\theta})\approx 3.73$, and $\gamma,\theta$
are the constants introduced in Section~\ref{se:useful-ineq}.
\end{lemma}

We postpone the proof of this lemma and 
proceed with the proof of the main theorem. 

\begin{Proof}(of Theorem \ref{main})
Let $\Gamma$ be a b.c.\ bilinear circuit for $\mathrm{conv}_n$, which
computes the linear forms $f_1,\dots,f_k$ on the first set of inputs.
Fix $1 \le r < n$, to be specified later, and set $R=\Rig{n-r}{f_1,\dots,f_k}$. 
By Lemma~\ref{lemma1} and Lemma~\ref{lemma2} there exists an $a \in \C^n$,
such that the following conditions hold:
\begin{enumerate}
\item $\max_{1\leq i\leq k} |f_i(a)|\leq 2\sqrt{\ln{(4k)}}\,R$,
\item $\compl{\Circ{a}}\geq\frac{1}{2}r\log{n} -c n$.
\end{enumerate}
By Lemma~\ref{lemma21} and the fact that $k\leq 3n^3$, we get
\begin{equation}
\label{maineq1}
  \size{\Gamma}+n\log{(2\sqrt{\ln{(12n^3)}}\, R)}\geq \compl{\Circ{a}}.
\end{equation}
On the other hand, the rigidity bound~(\ref{eq:rig-bound}) implies
the following upper bound on $R$ in terms of $\size{\Gamma}$:
\begin{equation*}
  \size{\Gamma}\geq \compl{f_1,\dots,f_k}\geq (n-r)\log{R} .
\end{equation*}
By combining this with~(\ref{maineq1}) and using the second condition above, we obtain
\begin{equation*}
 \Big(1+\frac{n}{n-r}\Big)\size{\Gamma} \geq\frac{r}{2}\log{n} - O(n\log\log{n}) .
\end{equation*}
Setting $\epsilon=r/n$ yields
$$
  \size{\Gamma}\geq \frac{\epsilon(1-\epsilon)}{2(2-\epsilon)}n\log{n}-O(n\log\log{n}).
$$
A simple calculation shows that the coefficient of the $n\log{n}$ term
attains the maximum $0.086$ for $\epsilon\approx 0.58$. 
Choosing $\epsilon=1/2$ for simplicity of exposition finishes the proof.
\end{Proof}  

Before going into the proof of Lemma~\ref{lemma2}, 
we provide a lemma on bounding the deviations of 
products of correlated normal random variables.

\begin{lemma}
\label{lemma51}
Let $Z=(Z_1,\dots,Z_r)$ be a centered Gaussian vector in $\C^r$. 
Define the complex covariance matrix of $Z$ by 
$\Sigma_r:=(\ew{Z_j \overline{Z}_k})_{j,k}$ and put 
$\delta := 2^{-(\gamma+\sqrt{2\theta})} \approx 0.02$. 
Then we have 
$\ew{|Z_1|^2\cdots |Z_r|^2} \geq \det{\Sigma_r}$ and 
$$
    \prob{|Z_1|^2\cdots |Z_r|^2 \geq \delta^r \det{\Sigma_r}}>\frac{1}{2} . 
$$
\end{lemma}

\begin{Proof}
For proving the bound on the expectation 
decompose $Z_r = \xi + \eta$ into a component $\xi$ in the span of 
$Z_1,\ldots,Z_{r-1}$ plus a  component $\eta$ orthogonal to 
this span in the Hilbert space of quadratic integrable 
random variables with respect to the inner product defined 
by the joint probability density of $Z$. Therefore, 
$|Z_r|^2 =|\xi|^2 +  \xi\overline{\eta} + \overline{\xi}\eta + |\eta|^2$,
hence by independence
\begin{eqnarray*}
  \ew{|Z_1|^2\cdots |Z_{r-1}|^2 |Z_{r}|^2} &=& 
 \ew{|Z_1|^2\cdots |Z_{r-1}|^2 |\xi|^2} + \ew{|Z_1|^2\cdots |Z_{r-1}|^2}\, \ew{|\eta|^2} \\ 
  & \ge & \ew{|Z_1|^2\cdots |Z_{r-1}|^2}\, \ew{|\eta|^2}.   
\end{eqnarray*}
By interpreting the Gramian determinant $\det\Sigma_r$ as the square volume of the 
parallelepiped spanned by the random vectors $Z_1,\ldots,Z_r$ in the Hilbert space, 
we obtain
$$
 \det\Sigma_r = \det\Sigma_{r-1}\, \ew{|\eta|^2}. 
$$
The desired bound on the expectation 
$\ew{|Z_1|^2\cdots |Z_{r}|^2} \ge \det\Sigma_r$
thus follows by induction on $r$. 
Noting that $\ew{|Z_r|^2}\ge\ew{|\eta|^2}$,
we also conclude from the above equation that 
\begin{equation}\label{eq:had}
 \ew{|Z_1|^2} \cdots \ew{|Z_{r}|^2} \ge \det\Sigma_r. 
\end{equation}

In order to prove the probability estimate for the random product
$|Z_1|^2\cdots |Z_{r}|^2$, 
we first transform the product into a sum by taking logarithms. 
For every $\epsilon >0$ Chebychev's inequality yields the bound
\begin{equation}
\label{eq51}
  \mathrm{P}\Big[\frac{1}{r}\Big|\sum_{j=1}^r(\log{|Z_j|^2-\mean{\log{|Z_j|^2}}})\Big|\geq
  \epsilon\Big]
  \leq \frac{\Var{\sum_{j=1}^r\log{|Z_j|^2}}}{\epsilon^2r^2}.
\end{equation}
For the variance we have by Lemma~\ref{prop22}
\begin{equation*}
\begin{split}
  \Var{\sum_{j=1}^r\log{|Z_j|^2}} =\sum_{j,k}\Cov{\log{|Z_j|^2},\log{|Z_k|^2}}\\
      \leq \sum_{j,k}\sqrt{\Var{\log{|Z_j|^2}}\Var{\log{|Z_k|^2}}} \leq r^2\theta .
\end{split}
\end{equation*}
Setting $\epsilon^2=2\theta$ in this equation and after exponentiating in~(\ref{eq51}) we obtain
\begin{equation}
\label{eq53}
  \prob{|Z_1|^2\cdots |Z_r|^2\leq 2^{-\epsilon r +\sum_{j=1}^r\mean{\log{|Z_j|^2}}}}\leq \frac{1}{2}.
\end{equation}
By combining the bound~(\ref{eq:had}) with Lemma~\ref{prop22} we get 
\begin{equation*}
  \log{\det{\Sigma_r}}\leq \sum_{i=1}^r\log{\mean{|Z_i|^2}}\leq \gamma r + \sum_{i=1}^r \mean{\log{|Z_i|^2}} .
\end{equation*}
Hence we conclude from~(\ref{eq53}) that 
\begin{equation*}
  \prob{|Z_1|^2\cdots |Z_r|^2 \le 2^{- (\epsilon +\gamma) r} \det{\Sigma_r}} \le \frac{1}{2},
\end{equation*}
from which the lemma follows.
\end{Proof}

\begin{Proof}(of Lemma \ref{lemma2})
By equation~(\ref{eq:ev-DFT}) we have $\lambda = \text{DFT}_n a$ and 
the singular values of the circulant~$\Circ{a}$ are given by the absolute values of the 
components of $\lambda$. Setting
$$
 \alpha = n^{-1/2}\lambda = n^{-1/2} \text{DFT}_n a ,
$$
we obtain for the $r$-mean square volume by (\ref{eq:prop31}) 
\begin{equation}
\label{eq421}
  \avol{r}^2(\Circ{a}) = n^r\sum_{|I|=r}\prod_{i\in I} |\alpha_{i}|^2 . 
\end{equation}

Now let $a$ be a standard Gaussian vector in the subspace $U$ of dimension~$r$. 
Let $W$ by the image of $U$ under the unitary transformation $n^{-1/2}\DFT$.
As a unitary transformation of $a$, $\alpha$ is standard Gaussian 
in the subspace $W$ (cf.~Section~\ref{sse:crv}).
This means that there 
is an orthonormal basis $b_1,\dots,b_{r}$ of $W$ such that
\begin{equation*}
  \alpha=\beta_1b_1+\dots+\beta_{r} b_{r},
\end{equation*}
where $(\beta_i)$ is standard Gaussian in $\C^{r}$.
Let $B\in\C^{n\times r}$  denote the matrix with the columns $b_1,\dots,b_{r}$
and let $B_I$ be the submatrix of $B$ consisting of the rows indexed by~$I$, 
for $I\subseteq [n]$ with $|I|=r$. 
Setting $\alpha_I=(\alpha_i)_{i\in I}$ we have $\alpha_I=B_I\beta$.
The complex covariance matrix of $\alpha_I$ is given by 
$\Sigma:= E[\alpha_I \alpha^\ast_I] = B_I B_I^\ast$, hence 
\begin{equation*}
  \det{\Sigma}= |\det{B_I}|^2.
\end{equation*}
We remark that $|\det{B_I}|^2$ can be interpreted as the volume contraction ratio 
of the projection $\C^n\to\C^I, \alpha\mapsto \alpha_I$ restricted to~$W$. 
For later purposes we also note that 
$\ew{|\alpha_i|^2} = \sum_j |B_{ij}|^2 \le 1$.

By the Binet-Cauchy formula and the orthogonality of the basis $(b_i)$ we get 
$$
\sum_{|I|=r}|\det{B_I}|^2=\det{(\langle b_i,b_j\rangle)_{1\leq i,j \leq r}}=1. 
$$ 
Therefore, we can choose an index set~$I$ such that  
\begin{equation*}
  |\det{B_I}|^2\geq \binom{n}{r}^{-1} \ge 2^{-n}.
\end{equation*}
By applying Lemma~\ref{lemma51} to the random vector $\alpha_I$ 
and using~(\ref{eq421}), we get that with probability at least $1/2$,
\begin{equation}\label{eq:le2q}
  \avol{r}^2(\Circ{a}) \geq n^r\delta^r \det{\Sigma} \geq n^r\delta^r 2^{-n},
\end{equation}
where $\delta=2^{-(\gamma+\sqrt{2\theta})}$. 
The mean square volume bound~(\ref{eq:msv-bound}) implies that 
$$
\compl{\Circ{a}} \geq \log{\avol{r}(\Circ{a}})- \frac{n}{2} 
            \geq \frac{1}{2} r \log{n}- \frac{1}{2}(2+\log\delta^{-1}) n ,
$$
with probability at least $1/2$.
This proves the lemma.    
\end{Proof}

\section{Multiplication and Division of Polynomials}
\label{se:appl}

By reducing the cyclic convolution to several other important computational problems, 
we are going to derive lower bounds of order $n\log n$ for these problems.
These bounds are optimal up to a constant factor. However, we did not attempt to optimize 
these factors. 

\subsection{Polynomial multiplication}

Let $f=\sum_{i=0}^{n-1}a_ix^i$ and $g=\sum_{i=0}^{n-1}b_ix^i$ be polynomials
in $\C[X]$ and $fg=\sum_{i=0}^{2n-2}c_ix^i$. 
Clearly, we can obtain the coefficients of
the cyclic convolution of~$f$ and~$g$ by adding $c_k$ to $c_{k+n}$ for 
$0\leq k <n$. 
This observation and Theorem~\ref{main} immediately imply the following corollary.

\begin{corollary}
\label{cor:poly-mult}
The bounded coefficient complexity of the multiplication of polynomials of degree less than $n$
is at least $\frac{1}{12}n\log{n}-O(n\log\log{n})$.
\end{corollary}
  
\subsection{Division with remainder}

We will first derive a lower bound on the inversion of power series
$\bmod\, X^{n+1}$ and then use this to get a lower bound for the division of polynomials. 

Let $\C[[X]]$ denote the ring of formal power series in the variable~$X$. 
We will study the problem to compute the first~$n$ coefficients 
$b_1,,\dots,b_{n}$ of the inverse in~$\C[[X]]$
$$
 f^{-1}= 1 + \sum_{k=1}^{\infty} b_k X^k
$$ 
of the polynomial $f= 1 - \sum_{i=1}^{n} a_i X^i$ 
given by  the coefficients $a_i$. 
We remark that the $b_k$ are polynomials in the $a_i$, which are recursively given by
\begin{equation*}
  b_0:=1,\quad b_k = \sum_{i=0}^{k-1}a_{k-i}b_i .
\end{equation*}
Note that the problem to invert power series is not bilinear. 
\cite{siev:72} and \cite{kung:74}
designed a b.c.\ circuit of size $O(n\log{n})$ 
solving this problem. 

We now prove a corresponding lower bound on the b.c.\ complexity of this problem by 
reducing polynomial multiplication to the problem to invert power series. 

\begin{theorem}
The map assigning to $a_1,,\dots,a_{n}$ the first $n$ coefficients 
$b_1,\dots,b_n$ of the inverse of $f = 1 - \sum_{i=1}^{n} a_i X^i$ in the ring of formal 
power series has bounded coefficient complexity greater than 
$\frac{1}{324} n\log{n}-O(n\log\log{n})$.
\end{theorem}

\begin{Proof}
Put $g=\sum_{i=1}^{n} a_i X^i$. 
The equation
\begin{equation*}
   1+\sum_{k=1}^{\infty}b_kX^k = \frac{1}{1-g} = \sum_{k=0}^{\infty}g^k .
\end{equation*}
shows that $g^2$ is the homogeneous quadratic part of $\sum_{k=1}^\infty b_k X^k$ in 
the variables~$a_i$. 

Let $\Gamma$ be an optimal b.c.\ circuit computing $b_1,\dots,b_n$. 
According to the proof in \cite[Theorem~7.1]{bucs:96}, there is a b.c.\ circuit 
of size at most~$9\,\size{\Gamma}$ 
computing the homogeneous quadratic parts of the $b_1,\ldots,b_n$ with respect to the 
variables $a_i$.  
This leads to a b.c.\ circuit of size at most $9\,\size{\Gamma}$ computing 
the coefficients of the squared polynomial $g^2$. 

Now let $m:=\lfloor n/3\rfloor$, and assume that 
$g=g_1+X^{2m}g_2$ with $g_1, g_2$ of degree smaller than $m$. Then 
\begin{equation*}
  g^2=g_1^2+2g_1g_2X^{2m}+g_2^2X^{4m},
\end{equation*}
By the assumption on the degrees we have no ``carries'' and we 
can therefore find the coefficients of the product polynomial $g_1g_2$ 
among the middle terms of $g^2$. 
Thus we obtain a b.c.\ circuit for the multiplication of polynomials of 
degree $m-1$. The theorem now follows from Corollary~\ref{cor:poly-mult}. 
\end{Proof}

We now show how to reduce the inversion of power series to the problem
of dividing polynomials with remainder. The reduction in the proof of
the following corollary is from \cite{stra:73-2}, see also
\cite[Section 2.5]{bucs:96}.

\begin{corollary}
Let $f,g$ be polynomials with $n=\deg{f}\geq m=\deg{g}$ and $g$ be monic. 
Let $q$ be the quotient and $r$ be the remainder of $f$ divided by $g$, so that
$f=qg+r$ and $\deg{r}<\deg{g}$. 
The map assigning to the coefficients of $f$ and $g$ the coefficients of 
the quotient~$q$ and the remainder~$r$ has bounded coefficient complexity 
at least $\frac{1}{324}n\log{n}-O(n\log\log{n})$.
\end{corollary}

\begin{Proof}
Dividing $f=X^{2n}$ by $g=\sum_{i=0}^n a_iX^{n-i}$, where $a_0=1$, we obtain: 
\begin{equation*} 
X^{2n}=\Big(\sum_{i=0}^{n}q_iX^i\Big)\Big(\sum_{i=0}^{n}a_iX^{n-i}\Big)+\sum_{i=0}^{n-1}r_iX^i.
\end{equation*}
By substituting $X$ with $1/X$ in the above equation and multiplying
with $X^{2n}$, we get 
\begin{equation*}
1=\Big(\sum_{i=0}^{n}q_iX^{n-i}\Big)\Big(\sum_{i=0}^na_iX^i\Big)+\sum_{i=0}^{n-1}r_iX^{2n-i}.
\end{equation*}
Since the remainder is now a multiple of $X^{n+1}$, we get
\begin{equation*}
\Big(\sum_{i=0}^{n}a_iX^i\Big)^{-1}\equiv
\Big(\sum_{i=0}^nq_iX^{n-i}\Big) \bmod X^{n+1}.
\end{equation*}
From this we see that the coefficients of the quotient are precisely
the coefficients of the inverse $\bmod \ X^{n+1}$ of 
$\sum_{i=0}^{n} a_iX^i$ in the ring of formal power series,
and the proof is finished.
\end{Proof}

\section{Unbounded scalar multiplications}
\label{se:help}

We extend our model of computation by allowing some instructions corresponding 
to scalar multiplications with constants of absolute value greater than two, 
briefly called {\em help gates} in the sequel. 
If there are at most $h$ help gates allowed, we denote the corresponding 
bounded coefficient complexity by the symbol $\mcal{C}_h$. 

We are going to show that our proof technique is robust in the sense that it 
still allows to prove $n\log{n}$ lower bounds if the number of help gates is restricted to 
$(1-\epsilon)n$ for fixed $\epsilon>0$.

\subsection{Extension of the mean square volume bound}

As a first step we extend the mean square volume bound~(\ref{eq:prop31}) and (\ref{eq:msv-bound}) 
for dealing with help gates. 

\begin{proposition}
\label{th:msv-help}
Assume $A\in \C^{m\times n}$ has the singular values 
$\sigma_1\ge\ldots\ge\sigma_p$, where $p:= \min{\{m,n\}}$. 
For all integers $s,h$ with $1\leq s \leq p - h$ we have  
\begin{equation*}
  \hcompl{h}{A} \geq \sum_{i=h+1}^{h+s}\log \sigma_{i}  - \frac{m}{2}  + h 
                \ge s \log \sigma_{h+s} - \frac{m}{2}  + h . 
\end{equation*}
\end{proposition}

\begin{Proof}
Let $\Gamma$ be a b.c.\ circuit with at most $h$ help gates, 
which computes the linear map corresponding to $A$. 
Without loss of generality, we may assume that $\Gamma$ has exactly $h$ 
help gates. 
Let $g_i$, $i\in I$, be the linear forms computed at the help gates of $\Gamma$. 
We transform the circuit $\Gamma$ into a b.c.\ 
circuit $\Gamma'$ by replacing each help gate with a
multiplication by zero. This new circuit is obviously a b.c.\ circuit
of size $\size{\Gamma'}=\size{\Gamma}-h$, computing a linear map 
corresponding to a matrix $B\in\C^{m\times n}$. 
The linear maps corresponding to $A$ and $B$ 
coincide on the orthogonal complement of $\mathrm{span}\{g_i \ | \ i\in I\}$ in $\C^m$, 
therefore $B=A+E$ for a matrix $E$ of rank at most~$h$.
From the perturbation inequality~(\ref{eq:pert}) we obtain that 
$$
  \sigma_{i}(B)\geq \sigma_{i+h}(A) \  \mbox{ for $i \le p-h$.} 
$$
By (\ref{eq:prop31}) this implies for $s\le p-h$ that 
$$
 \avol{s}^2 (B)\ \ge\ \sum_{0 < i_1 < \cdots < i_s\le p-h} \sigma^2_{i_1}(B)\cdots \sigma^2_{i_s}(B)
               \ \ge\ \sum_{h < i_1 < \cdots <i_s\le p} \sigma^2_{i_1}(A)\cdots \sigma^2_{i_s}(A).
$$
On the other hand, by the mean square volume bound~(\ref{eq:msv-bound}) we have 
$$
 \size{\Gamma}-h=\size{\Gamma'} \geq \log\avol{s}(B) - \frac{m}{2} .
$$
Combining the last two estimates completes the proof. 
\end{Proof}

\begin{remark}
\begin{description}

\item[\rm 1] Proposition~\ref{th:msv-help} implies that 
$\hcompl{(1-\epsilon)n}{\mathrm{DFT}_n} \ge \epsilon (\frac{1}{2}n\log{n} - n)$
for the Discrete Fourier Transform $\mathrm{DFT}_n$, 
provided $0<\epsilon\le 1$.  

\item[\rm 2] Note that the number $h$ of help gates  
may be replaced by the dimension of the subspace spanned by the 
linear functions computed at the help gates. 

\item[\rm 3] Proposition~\ref{th:msv-help} can be seen as a variant 
of the spectral lemma in \cite{chaz:98}.
Using entropy considerations, Chazelle obtained the slightly worse lower bound 
$\Omega((r-2h)\log{\sigma_r})$ for the b.c.\ complexity
of a matrix $A\in\R^{n\times n}$  with at most $h$ help gates.  
While this allows to handle at most $n/2$ help gates,  Chazelle's result is stronger 
in the sense that it involves a more general notion of help gates, 
which are allowed to compute {\em any} function of the previous intermediate results. 

\end{description}
\end{remark}

\subsection{Extremal values of Gaussian random vectors}

In this section we derive the following auxiliary result about the distribution
of the maximal absolute value of the components of a Gaussian random vector. 

\begin{lemma}\label{le:extr}
\begin{description}
\item[1] A centered Gaussian random vector $X=(X_1,\ldots,X_n)$ in $\R^n$ 
with $\max_i \ew{X_i^2} \le 1$ satisfies for any $\epsilon>0$ 
$$
 \lim_{n\to\infty}\prob{ \max_i |X_i| > \sqrt{2\ln n} + \epsilon } = 0.
$$
\item[2] A centered Gaussian random vector $(Z_1,\ldots,Z_n)$ in~$\C^n$ with 
$\max_i\ew{|Z_i|^2} \le 1$ satisfies for any $\epsilon>0$ 
$$
 \lim_{n\to\infty}\prob{ \max_i |Z_i| > 2\sqrt{\ln (2n)} + \epsilon } = 0.
$$
\end{description}
\end{lemma}

\begin{proof}
1. Since $X$ is centered we have for any $u\in\R$
$$
\prob{\max_i |X_i|\ge u}  \le \prob{\max_i X_i\ge u} + \prob{\max_i (-X_i)\ge u} 
                          \le 2 \prob{\max_i X_i \ge u} .
$$
For proving the first assertion it is therefore sufficient to show that for 
any $\epsilon>0$ 
\begin{equation}\label{eq:maxd}
 \lim_{n\to\infty}\prob{ \max_i X_i > \sqrt{2\ln n} + \epsilon } = 0.
\end{equation}

For this we may assume that the components of $X$ are uncorrelated.
In fact, Slepian's inequality (see~\cite{leta:91}) implies that for 
centered Gaussian vectors $X=(X_1,\ldots,X_n)$ and $Y=(Y_1,\ldots,Y_n)$ we have 
$$
 \prob{\max_i X_i \le u} \le  \prob{\max_i Y_i \le u}  
$$
provided $\ew{X_i^2}=\ew{Y_i^2}$ and $\ew{X_i X_j}\le\ew{Y_i Y_j}$ for all $i,j$. 

We may also assume that all the $X_i$ have variance~$1$ since 
the distribution function 
$$
 F_\sigma(u):= \frac{1}{\sigma\sqrt{2\pi}}\int_\infty^u \exp(-\frac{t^2}{2\sigma^2})dt .
$$
of a centered normal random variable with variance $\sigma^2\le 1$ 
satisfies $F_1(u) \le F_\sigma(u)$ for all $u\ge 0$. 
Hence, if $X$ is a Gaussian vector with uncorrelated components $X_i$ of 
variance $\sigma_i^2\le 1$, we have
$$
  F_{1}(u)^n \le \prod_{i=1}^n F_{\sigma_i}(u) = \prob{\max_i X_i \le u} .  
$$

In the case where $X_1,\ldots,X_n$ are independent and standard normal distributed
we have according to~\cite{cram:46} that 
$$
 \ew{\max_i X_i} = \sqrt{2\ln n} + o(1),\quad
 \Var{\max_i X_i} = \frac{\pi^2}{12}\,\frac{1}{\ln n}\,(1 +o(1)),\quad n\to\infty 
$$
and Claim~(\ref{eq:maxd}) follows from Chebychev's inequality.

2. The second assertion follows from the first one applied to the 
Gaussian vector~$W$ with values in $\R^{2n}$ given by the real and imaginary parts 
of the $Z_i$ (in some order). Note that 
$\max_{1\le i\le n} |Z_i| \le \sqrt{2}\max_{1\le j\le 2n} |W_j|$. 
\end{proof}

\subsection{Cyclic convolution and help gates}

Our goal is to prove the following extension of Theorem~\ref{main}.

\begin{theorem}
\label{main-ext}
The bounded coefficient complexity
with at most $(1-\epsilon)n$ help gates of the $n$-dimensional cyclic convolution 
$\mathrm{conv}_n$ is at least $\Omega(n\log{n})$
for fixed $0 < \epsilon \le 1$. 
\end{theorem}

The proof follows the same line of argumentation as in Section~\ref{se:conv}.
We first state and prove an extension of Lemma~\ref{lemma2}.

\begin{lemma}
\label{lemma2-ext}
Let $U\subseteq \C^n$ be a subspace of dimension $r$ and $h\in\N$ with $h < r$. 
For a standard Gaussian vector~$a$ in $U$, we have
\begin{equation*}
  \prob{\hcompl{h}{\Circ{a}}\geq \frac{1}{2}(r-h)\log{n} - n(c+\log\log n)} >\frac{1}{2},
\end{equation*}
for some constant $c>0$. 
\end{lemma}

\begin{proof} 
As in the proof of Lemma~\ref{lemma2} we  assume that the random vector
$\alpha = n^{-1/2} \text{DFT}_n a$ 
is standard Gaussian with values in some $r$-dimensional subspace~$W$. 
Recall that $\sqrt{n}\,|\alpha_i|$ are the singular values of $\Circ{a}$.  
We denote by
$|\alpha^{(1)}| \ge \ldots \ge |\alpha^{(n)}|$ 
the components of $\alpha$ with decreasing absolute values.
In particular, $|\alpha^{(1)}| = \max_i |\alpha^{(i)}|$. 
Proposition~\ref{th:msv-help} implies that 
\begin{eqnarray*}
  \hcompl{h}{\Circ{a}} &\geq& 
      \sum_{i=h+1}^{r}\log (\sqrt{n}\,|\alpha^{(i)}|)  - \frac{n}{2} + h \\ 
  &=& \frac{1}{2}(r-h)\log{n} + \log\bigg(\prod_{i=h+1}^{r} |\alpha^{(i)}|\bigg) - \frac{n}{2} + h. 
\end{eqnarray*}

In the proof of Lemma~\ref{lemma2}~(\ref{eq:le2q}) we showed that 
$\avol{r}^2(\Circ{a}) \ge n^r\delta^r 2^{-n}$ 
with probability at least $1/2$. In the same way, one can show that 
with probability at least $3/4$ we have 
$\avol{r}^2(\Circ{a}) \ge n^r c_1^{n}$
for some fixed constant $c_1>0$.  
From the estimate
$$ 
\sum_{|I|=r}\ \prod_{i\in I} |\alpha_i|^2 \ \le \ 2^n \prod_{i=1}^r |\alpha^{(i)}|^2
$$
we thus obtain that 
$\prod_{i=1}^r |\alpha^{(i)}|^2 \ge (c_1/2)^n$ 
with  probability at least $3/4$. 

By applying Lemma~\ref{le:extr} to the centered Gaussian random variable $\alpha$ 
we obtain that with probability at least $3/4$
$$
  \max_i |\alpha^{(i)}|^2 = |\alpha^{(1)}|^2 \le c_2 \log n  
$$
for some fixed constant $c_2>0$. 
(Recall that $\ew{|\alpha^{(i)}|^2}\le 1$.) 

Altogether, we obtain that with probability at least $1/2$ we have 
$$
  \prod_{i=h+1}^r |\alpha^{(i)}|^2 \ \ge\  \frac{\prod_{i=1}^r |\alpha^{(i)}|^2}{|\alpha^{(1)}|^{2h}}
    \ \ge\  \bigg(\frac{c_1}{2 c_2\log n}\bigg)^n .  
$$
This completes the proof of the lemma. 
\end{proof}

\begin{proof}(of Theorem~\ref{main-ext}) 
Let $\Gamma$ be a b.c.\ bilinear circuit computing $\mathrm{conv}_n$ 
using at most $h\le (1-\epsilon)n$ help gates, $0< \epsilon \le 1$.  
Referring to the partition of instructions in Definition~\ref{def:circuit}, 
we assume that $\Gamma^{(1)}$ uses $h_1$ help gates, 
and that $\Gamma^{(2)},\Gamma^{(3)},\Gamma^{(4)}$ use a total of $h_2$ help gates. 
Thus $h_1 + h_2 = h$. 
Let $f_1,\dots,f_k$ denote the linear forms computed by $\Gamma^{(1)}$.

Assume $h_2 < r < n-h_1$ and set $R=\Rig{n-r}{f_1,\dots,f_k}$. 
By Lemma~\ref{lemma1} and Lemma~\ref{lemma2-ext} there exists an $a \in \C^n$,
such that the following conditions hold:
\begin{enumerate}
\item $\max_{1\leq i\leq k} \log |f_i(a)|\leq \log(2\sqrt{\ln{(4k)}}\,R) \le \log R + O(\log\log n)$,
\item $\hcompl{h_2}{\Circ{a}}\geq\frac{1}{2}(r-h_2)\log{n} - O(n\log\log n)$.
\end{enumerate} 
On the other hand, by Proposition~\ref{th:msv-help}
and using $\sigma_{n-r}(f_1,\ldots,f_k) \ge R$, we get
\begin{equation*}
  \size{\Gamma}\geq \hcompl{h_1}{f_1,\dots,f_k}\geq (n-r-h_1)\log{R} - \frac{k}{2} .
\end{equation*}
The proof of Lemma~\ref{lemma21} shows that 
$$
  \size{\Gamma}+n\max_{1\leq i\leq k} \log |f_i(a)| \ \geq\ \hcompl{h_2}{\Circ{a}}.
$$
By combining all this we obtain
\begin{equation*}
 \Big(1+\frac{n}{n-r-h_1}\Big)\size{\Gamma} + \frac{nk}{2(n-r-h_1)} + O(n\log\log{n}) 
   \geq \frac{1}{2}(r-h_2)\log{n} . 
\end{equation*}
We set now $r:=\lfloor(h_2 + n-h_1)/2\rfloor$. 
Then $r+h_1 \le (1 - \frac{\epsilon}{2}) n$ and $r-h_2 \ge \frac{\epsilon}{2}\, n -1$.  
By plugging this into the above inequality we obtain
$$
  \frac{\epsilon +2}{\epsilon}\,\size{\Gamma} + \frac{k}{\epsilon} + O(n\log\log{n}) 
     \geq \frac{\epsilon}{4}\, n\log{n} . 
$$
Let $\kappa := \frac{\epsilon^2}{8}$. 
If $k \le \kappa n\log{n} + n$, then 
$\size{\Gamma}\ge \frac{\epsilon^2}{8(\epsilon+2)}\, n\log{n} - O(n\log\log{n})$. 
On the other hand, if $k >  \kappa n\log{n} +n$, then trivially
$$
 \size{\Gamma} \ge  \hcompl{h_1}{f_1,\dots,f_k} \geq k - n \ge \kappa n\log{n} . 
$$
This completes the proof of the theorem. 
\end{proof}

\begin{acks}
We are grateful to Joachim von zur Gathen for bringing the paper~\cite{rraz:02} to our attention. 
We thank Satyanarayana Lokam for suggesting to extend our lower bounds involving help gates 
from the linear to the bilinear case. We also thank Tom Schmitz and Mario Wschebor 
for useful discussions about probability.
This work has been supported by 
Forschungspreis 2002 der Universit\"at Paderborn 
and by the Paderborn Institute for Scientific Computation (PaSCo).  
\end{acks}
 
\bibliographystyle{acmtrans}
\bibliography{/fields/user/pbuerg/mypapers/biblio/lit-bank}

\begin{thebibliography}{}

\bibitem[\protect\citeauthoryear{Bellman}{Bellman}{1997}]{bell:97}
{\sc Bellman, R.} 1997.
\newblock {\em Introduction to matrix analysis}.
\newblock SIAM, Philadelphia, PA.

\bibitem[\protect\citeauthoryear{B{\"u}rgisser, Clausen, and
  Shokrollahi}{B{\"u}rgisser et~al\mbox{.}}{1997}]{bucs:96}
{\sc B{\"u}rgisser, P.}, {\sc Clausen, M.}, {\sc and} {\sc Shokrollahi, M.}
  1997.
\newblock {\em Algebraic Complexity Theory}. {G}rundlehren der mathematischen
  {W}issenschaften, vol. 315.
\newblock Springer Verlag.

\bibitem[\protect\citeauthoryear{Chazelle}{Chazelle}{1998}]{chaz:98}
{\sc Chazelle, B.} 1998.
\newblock A spectral approach to lower bounds with applications to geometric
  searching.
\newblock {\em SIAM Journal on Computing\/}~{\em 27(2)}, 545--556.

\bibitem[\protect\citeauthoryear{Courant and Hilbert}{Courant and
  Hilbert}{1931}]{couh:31}
{\sc Courant, R.} {\sc and} {\sc Hilbert, D.} 1931.
\newblock {\em Methoden der mathematischen {P}hysik. {I}}.
\newblock Springer-Verlag, Berlin.
\newblock Zweite Auflage.

\bibitem[\protect\citeauthoryear{Cram\'er}{Cram\'er}{1946}]{cram:46}
{\sc Cram\'er, H.} 1946.
\newblock {\em Mathematical Methods of Statistics}. Princeton Mathematical
  Series, vol.~9.
\newblock Princeton University Press.

\bibitem[\protect\citeauthoryear{Golub and Van~Loan}{Golub and
  Van~Loan}{1996}]{govl:96}
{\sc Golub, G.~H.} {\sc and} {\sc Van~Loan, C.} 1996.
\newblock {\em Matrix Computations}.
\newblock The John Hopkins University Press, Baltimore.

\bibitem[\protect\citeauthoryear{Kung}{Kung}{1974}]{kung:74}
{\sc Kung, H.} 1974.
\newblock On computing reciprocals of power series.
\newblock {\em Num.~Math.\/}~{\em 22}, 341--348.

\bibitem[\protect\citeauthoryear{Lang}{Lang}{1984}]{lang:84}
{\sc Lang, S.} 1984.
\newblock {\em Algebra\/}, Second ed.
\newblock Addison-Wesley.

\bibitem[\protect\citeauthoryear{Ledoux and Talagrand}{Ledoux and
  Talagrand}{1991}]{leta:91}
{\sc Ledoux, M.} {\sc and} {\sc Talagrand, M.} 1991.
\newblock {\em Probability in {B}anach Spaces}. Ergebnisse der Mathematik und
  ihrer Grenzgebiete, 3.\ Folge, vol.~23.
\newblock Springer Verlag.

\bibitem[\protect\citeauthoryear{Lokam}{Lokam}{1995}]{loka:95}
{\sc Lokam, S.} 1995.
\newblock Spectral methods for matrix rigidity with applications to size-depth
  tradeoffs and communication complexity.
\newblock In {\em Proc.~36th FOCS}. 6--15.

\bibitem[\protect\citeauthoryear{Morgenstern}{Morgenstern}{1973}]{morg:73}
{\sc Morgenstern, J.} 1973.
\newblock Note on a lower bound of the linear complexity of the fast {F}ourier
  transform.
\newblock {\em J.~ACM\/}~{\em 20}, 305--306.

\bibitem[\protect\citeauthoryear{Morgenstern}{Morgenstern}{1975}]{morg:75}
{\sc Morgenstern, J.} 1975.
\newblock The linear complexity of computation.
\newblock {\em J.~ACM\/}~{\em 22}, 184--194.

\bibitem[\protect\citeauthoryear{Nisan and Wigderson}{Nisan and
  Wigderson}{1995}]{niwi:95}
{\sc Nisan, N.} {\sc and} {\sc Wigderson, A.} 1995.
\newblock On the complexity of bilinear forms.
\newblock In {\em Proc. of the 27th ACM Symposium on the Theory of Computing}.
  723--732.

\bibitem[\protect\citeauthoryear{Pudl\'ak}{Pudl\'ak}{1998}]{pudl:98}
{\sc Pudl\'ak, P.} 1998.
\newblock A note on the use of determinant for proving lower bounds on the size
  of linear circuits.
\newblock {\em ECCC Report\/}~{\em 42}.

\bibitem[\protect\citeauthoryear{Raz}{Raz}{2002}]{rraz:02}
{\sc Raz, R.} 2002.
\newblock On the complexity of matrix product.
\newblock In {\em Proc.~34th STOC}. 144--151.
\newblock Also available as ECCC Report 12, 2002.

\bibitem[\protect\citeauthoryear{Sieveking}{Sieveking}{1972}]{siev:72}
{\sc Sieveking, M.} 1972.
\newblock An algorithm for division of power series.
\newblock {\em Computing\/}~{\em 10}, 153--156.

\bibitem[\protect\citeauthoryear{Strassen}{Strassen}{1973a}]{stra:73-2}
{\sc Strassen, V.} 1973a.
\newblock Die {Berechnungskomplexit\"at} von elementar\-sym\-metrischen
  {Funktionen} und von {Interpolationskoeffizienten}.
\newblock {\em Num.~Math.\/}~{\em 20}, 238--251.

\bibitem[\protect\citeauthoryear{Strassen}{Strassen}{1973b}]{stra:73-1}
{\sc Strassen, V.} 1973b.
\newblock {Vermeidung} von {Divisionen}.
\newblock {\em Crelles J.~Reine Angew.\ Math.\/}~{\em 264}, 184--202.

\bibitem[\protect\citeauthoryear{Valiant}{Valiant}{1976}]{vali:76-2}
{\sc Valiant, L.} 1976.
\newblock Graph theoretic properties in computational complexity.
\newblock {\em J.~Comp. Syst. Sci.\/}~{\em 13}, 278--285.

\bibitem[\protect\citeauthoryear{Valiant}{Valiant}{1977}]{vali:77}
{\sc Valiant, L.} 1977.
\newblock Graph theoretic arguments in low-level complexity.
\newblock Number~53 in LNCS. Springer Verlag, 162--176.

\end{thebibliography}

\end{document}